*Article*

# Reconstruction of Optical Coherence Tomography Images from Wavelength-space Using Deep-learning

Maryam Viqar [1,2,*], Erdem Sahin [1], Elena Stoykova [2] and Violeta Madjarova [2]

1. Faculty of Information Technology and Communication Sciences, Tampere University, 33720, Tampere, Finland; maryamviqar92@gmail.com (M.V.); erdem.sahin@tuni.fi (E.R.);
2. Institute of Optical Materials and Technologies, Bulgarian Academy of Sciences, Sofia, Bulgaria; vmadjarova@iomt.bas.bg (V.M.); elena.stoykova@gmail.com (E.S.);
* Correspondence: maryamviqar92@gmail.com (M.V.);

**Abstract:** Conventional Fourier-domain Optical Coherence Tomography (FD-OCT) systems depend on resampling into wavenumber ($k$) domain to extract the depth profile. This either necessitates additional hardware resources or amplifies the existing computational complexity. Moreover, the OCT images also suffer from speckle noise, due to systemic reliance on low coherence interferometry. We propose a streamlined and computationally efficient approach based on Deep-Learning (DL) which enables reconstructing speckle-reduced OCT images directly from the wavelength ($\lambda$) domain. For reconstruction, two encoder-decoder styled networks namely Spatial Domain Convolution Neural Network (SD-CNN) and Fourier Domain CNN (FD-CNN) are used sequentially. The SD-CNN exploits the highly degraded images obtained by Fourier transforming the ($\lambda$) domain fringes to reconstruct the deteriorated morphological structures along with suppression of unwanted noise. The FD-CNN leverages this output to enhance the image quality further by optimization in Fourier domain (FD). We quantitatively and visually demonstrate the efficacy of the method in obtaining high-quality OCT images. Furthermore, we illustrate the computational complexity reduction by harnessing the power of DL models. We believe that this work lays the framework for further innovations in the realm of OCT image reconstruction.

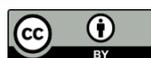

## 1. Introduction

Recent advancements in machine vision expedited by Deep-Learning (DL) methods have revolutionized the fields of image enhancement, reconstruction, classification, or feature extraction with enormous applications in fields of bio-medical or biological imaging [1] and healthcare [2]. The benchmark results have motivated researchers to use DL models like Convolution Neural Networks (CNNs) in several imaging modalities like Magnetic Resonance Imaging (MRI), Computed Tomography (CT), Optical Coherence Tomography (OCT), etc. to reconstruct not only visually pleasant high-quality images but to further leverage medically-significant and elucidative scans with reduced computational complexity. OCT is one such three-dimensional (3D) imaging technique based on interference phenomenon. It is a high-speed, non-invasive, and highly sensitive imaging modality. Nonetheless, certain inherent characteristics in the fundamental operation of OCT impose constraints on its imaging capabilities including resolution, depth penetration, signal-to-noise ratio (SNR) and time-complexity. Amongst Fourier Domain (FD) types of the OCT systems, the Swept Source (SS) is an advanced OCT technology based on the sweeping laser source that enables real-time, 3D biomedical image acquisition at very high speeds [3]-[5]. The imaging performance is directly influenced by factors such as the sweep rate and range, and the instantaneous linewidth of the laser source. A common drawback in most FD-OCT systems, either swept source or spectral domain, is the necessity to have data linear in the wavenumber domain for the acquired OCT interference fringes before subjecting them to the inverse Fourier transformation.

The widely used approach to address the problem of calibration and linearization in wavenumber domain is to perform resampling by employing interpolation techniques [6]. Some methods utilize resampling methods employing linear [7] or cubic spline [8] interpolation, or Kaiser-Bessel window function [9]. Another way is to perform reconstruction using techniques like Non-uniform Discrete Fourier Transform (DFT) [10], which can work with non-uniformly spaced data. Such techniques can potentially cause repercussions such as noise, artifacts, time complexity affecting the imaging rate, etc. Additionally, these methods require a remapping function estimated via a separate calibration procedure and could potentially necessitate extra hardware.

Calibration and k-mapping are challenging issues for the contemporary high-speed systems like "MHz-OCT" or "multi-MHz-OCT" systems. These systems are categorized using their A-scan rates, and they use wavelength swept lasers as one of benchmark contributors [11]. Several types of swept lasers are used, namely short cavity lasers [12], stretched pulse lasers [13], MEMS-VCSELs [14], Fourier domain mode locked lasers (FDMLs) [15], to name a few. For all these ultrahigh A-scan rates, both resampling and calibration stand as vital data-processing prerequisites. Recently, diverse alternatives have been employed by scholars using hardware-based techniques in place of aforementioned resampling and calibration approaches. To linearly sample the OCT fringes with respect to the wavenumber domain, the OCT system was realized with an optical k-clock [16]. However, this approach exhibited reliability issues when operating beyond 1.3 Giga Samples per second (GSPS) to achieve multi-MHz speeds per A-scan. And the system was vulnerable to clocking glitches and inaccuracies in the sampling process. A more recent work [17] performs calibration and resampling between consecutive sweeps by utilizing dual-channel acquisition of the OCT signal. This approach effectively overcomes the intrinsic limitations of optical clocking, but at the same time it increases the complexity. Given that the MEMS based systems can encounter fluctuations in wavelength versus time over several sweeps, calibration is conducted for each A-scan, thus significantly increasing the computational burden. Akinetic lasers have been used by H. Lee et. al. [18] to linearize the wavenumber domain, but their high cost makes them unsuitable for commercialization. The progress done in the imaging rates has been steered primarily by the innovations in the fundamental OCT hardware. In a quest to improve the performance, these benchmark OCT systems could further be advanced by extracting more potential from the software using revolutionizing deep-learning techniques.

In addition to wavenumber linearization issue, the image quality in OCT reconstruction is also degraded by the inevitable speckle due to coherent nature of the light source. The speckle-related artifacts can be suppressed by averaging successive B-scans when the morphology of the sample remains the same with these scans [19]. However, these iterations can be computationally complex and may cause blurring of the morphology variations when excessive averaging is done. Moreover, the dynamic nature of live samples, where movements occur, make these methods ineffective and may introduce degradations in the images.

The hardware-based resampling and calibration methods demand high-performance supplementary components, whereas software-based approaches can be established using the already built-in system with minor alterations. Thus, instead of addressing the constraints of the aforementioned hardware methods, the OCT image reconstruction can be performed using the DL techniques. The DL based reconstruction methods have gained significant attention in various imaging modalities like MRI [20]-[22], CT [23, 24], ultrasonography [25, 26] and many others, compared to methods based on hand-crafted feature extractors. As far as OCT is concerned, several attempts have been made on reconstruction of high-quality B-scans from the wavenumber domain data, e.g. reconstruction from under sampled wavenumber-domain spectrum [27]-[29].

In this work, we propose a DL framework to reconstruct high-quality OCT images directly from the wavelength-domain. We take into consideration the Fourier-spatial relation between the power spectral density of the interreference signal and autocorrelation of optical path length differences (Weiner- Khinchin theorem). This physical prior is used to guide the data-driven CNN based network in reconstruction of OCT B-scans. True to our knowledge, the wavelength-domain based image reconstruction problem using a dual DL-framework in the spatial and Fourier domains has been untouched so far. Utilizing the Fourier-space information can further enhance the current reconstruction quality along with reduced time-complexity compared to the images obtained from commercial systems [30].

The main highlights of this work are as follows:

- We propose a DL-framework with two networks implemented sequentially to reconstruct images. One network optimizes the $\lambda$-domain interference spectrum (or non-linear wavenumber-domain) in Fourier space and is called FD-CNN model. The other network, called SD-CNN, optimizes the spatial domain. This dual optimization strategy helps DL-framework to extract the non-linear relationships in two different domains leading to more robust information extraction.
- Each DL-network used is an encoder-decoder architecture based on UNET [31]. Unlike the original UNET network, the model in this work also incorporates the residual connections and attention for enhanced performance. The FD-CNN uses frequency-loss [32] function to account for missing linearity in wavenumber domain. The combination of two optimization models, facilitates the performance by guiding the dual-domain data-driven network. The experimental results show that this architecture is more streamlined and capable in generating OCT images efficiently.
- We also embedded a layer containing wavenumbers corresponding to each pixel value axially for every OCT A-scan as input matrix, to further guide the network training in SD-CNN. The ground truth is 7-averaged B-scans, obtained from the commercial OCT system [30] to suppress speckle noise.
- We observe the computational time-complexity and image enhancement characteristics of the proposed model in terms of morphological details, contours, edges (high-frequency content), and suppression of unwanted speckle noise by performing comparative analysis. For a fair analysis, comparison is performed between the proposed reconstruction method and the processing approach in the commercial OCT Optores GmbH system [30].

The remainder of this paper is organized as follows: Section 2 Materials and Methods focuses on the problem formulation followed by the proposed method and materials used, Section 3 presents the performed experiments, results, comparisons and analysis, Section 4 provides the discussion on the proposed work.

## 2. Materials and Methods

### 2.1. Problem formulation

The light reflectivity information in a SS-OCT is detected using a Michelson interferometer where the light source sweeps the spectrum linearly in time [33]. Here, the intensity $I_D$ of the acquired interferometric pattern can be expressed mathematically as:

$$I_D = I_D(k(t)) = I_1 + I_2 + I_3 + N_G(t), \tag{1}$$

where, $I_1 = S(k(t)) \left\{ \frac{\rho}{4} [r_r + r_{S1} + r_{S2} + \cdots] \right\}$,

$$I_2 = S(k(t))\left\{\frac{\rho}{2}[\sum_{i=1}^{\infty}\sqrt{r_r r_{Si}}(\cos(k(t)(d_r - d_{Si})))]\right\},$$

$$I_3 = S(k(t))\left\{\frac{\rho}{2}[\sum_{n \neq m}^{\infty}\sqrt{r_{Sj} r_{Si}}(\cos(k(t)(d_{Sj} - d_{Si})))]\right\}.$$

In Eq. (1), S(k(t)) represents the spectral shape of the light source that is wavenumber (k) dependent, ρ is the responsivity of the detector, $r_r$ and $r_{Si}$ (or $r_{Sj}$) represent reflectivity from the reference mirror (at depth $d_r$) and i$^{th}$ (or j$^{th}$) sample particle at depth $d_{Si}$ (or at $d_{Sj}$), respectively. $N_G$ is the additive white Gaussian noise term created by various sources. Furthermore, the term $I_1$ is the DC component and is removed using biasing technique. The term $I_2$ dominates over $I_3$, as the reflectivity of the reference is higher compared to reflectivity of the object particles. Then only the term $I_2$ is kept in Eq. (1) that contains reflectivity information for particles at varying depth within the sample. This leads to further simplification:

$$I_D = S(k(t))\left\{\frac{\rho}{2}\sum_{i=1}^{\infty}\sqrt{r_r r_{Si}}(\cos(k(t)(d_r - d_{Si})))\right\} + N_G(t), \tag{2}$$

The reflectivity distribution $r_{Si}$ of the sample particles can be extracted by performing IDFT (Inverse Discrete Fourier Transform) in the wavenumber space (k) as:

$$i_z = \mathcal{F}^{-1}\{I_D\}, \tag{3}$$

$$i_z = \frac{1}{2}\sum_{i \neq 1}^{\infty}\sqrt{r_r r_{Si}}(\alpha(d_r - d_{Si}) + \alpha(-(d_r - d_{Si}))). \tag{4}$$

Here $\mathcal{F}^{-1}$ is the symbolic representation of IDFT in Eq. (3) and $\alpha(d)$ in Eq. (4) comes from IDFT of the *S(k)*. As the IDFT is mathematically based on the principle of uniformly spaced data-points (wavenumbers), it is critical to sample the interferometric signal uniformly in the wavenumber domain with identical step size. This contrasts with the spectrum acquired, as this spectrum is uniformly distributed in wavelength domain (spans over time interval -Δt/2 to Δt/2), having a linear relationship with time expressed as:

$$\lambda = \beta t + \lambda_0, \tag{5}$$

where $\beta$ represents the sweeping speed of the source, $\lambda$ represents wavelength that varies from $\lambda_{min}$ to $\lambda_{max}$ and $\lambda_0$- is the central wavelength in Eq. (5). Now if we use $\lambda = 2\pi/k$, we can write Eq. (6) as:

$$t = \frac{1}{\beta}(\lambda - \lambda_0) = \frac{2\pi}{\beta}\left(\frac{1}{k} - \frac{1}{k_0}\right). \tag{6}$$

Upon expansion using power series we get:

$$t = \frac{2\pi}{\beta}\left[-\frac{1}{k_0}\left(\frac{k}{k_0} - 1\right) + \frac{1}{k_0}\left(\frac{k}{k_0} - 1\right)^2 - \frac{1}{k_0}\left(\frac{k}{k_0} - 1\right)^3 + \cdots\right], \tag{7}$$

**k** and $k_0$ are wavenumbers corresponding to the λ and $\lambda_0$ respectively. Moreover, for the recent advanced light sources like FDML [15], there exists a sinusoidal relationship even between wavelength and time, which can be written as:

$$\lambda(t) = \lambda_0 + \frac{\Delta\lambda}{2}(\sin(2\pi f_t t)), \tag{8}$$

where, Δλ is the bandwidth of the light source, $f_t$ is the tunning frequency of Fabry–Perot (FP) filter in FDML source.

The crucial feature of an OCT system is the depth encoding spectrum, which must undergo IDFT (Inverse Discrete Fourier Transform) to retrieve the depth information as obtained in Eq. (4). It is evident, from Eq. (7) and Eq. (8), that the spectrum is a non-linear function of the wavenumbers, and that the non-linear terms in expansion cannot be neglected hence the spectrum needs to go through calibration and k-linearization processes, to extract B-scans [34].

Another commonly encountered problem in OCT scans is speckle noise. Speckle is an evitable artifact due to coherent nature of light source used in OCT systems. It directly results from unwanted interference of scattered light from different points within the sample volume. It can affect the quality of images and hinder the quantitative analysis severely. It leads to loss of morphological details by affecting the contrast. To reconstruct good quality OCT images, it is important to reduce the speckle noise adequately without the loss of structural details.

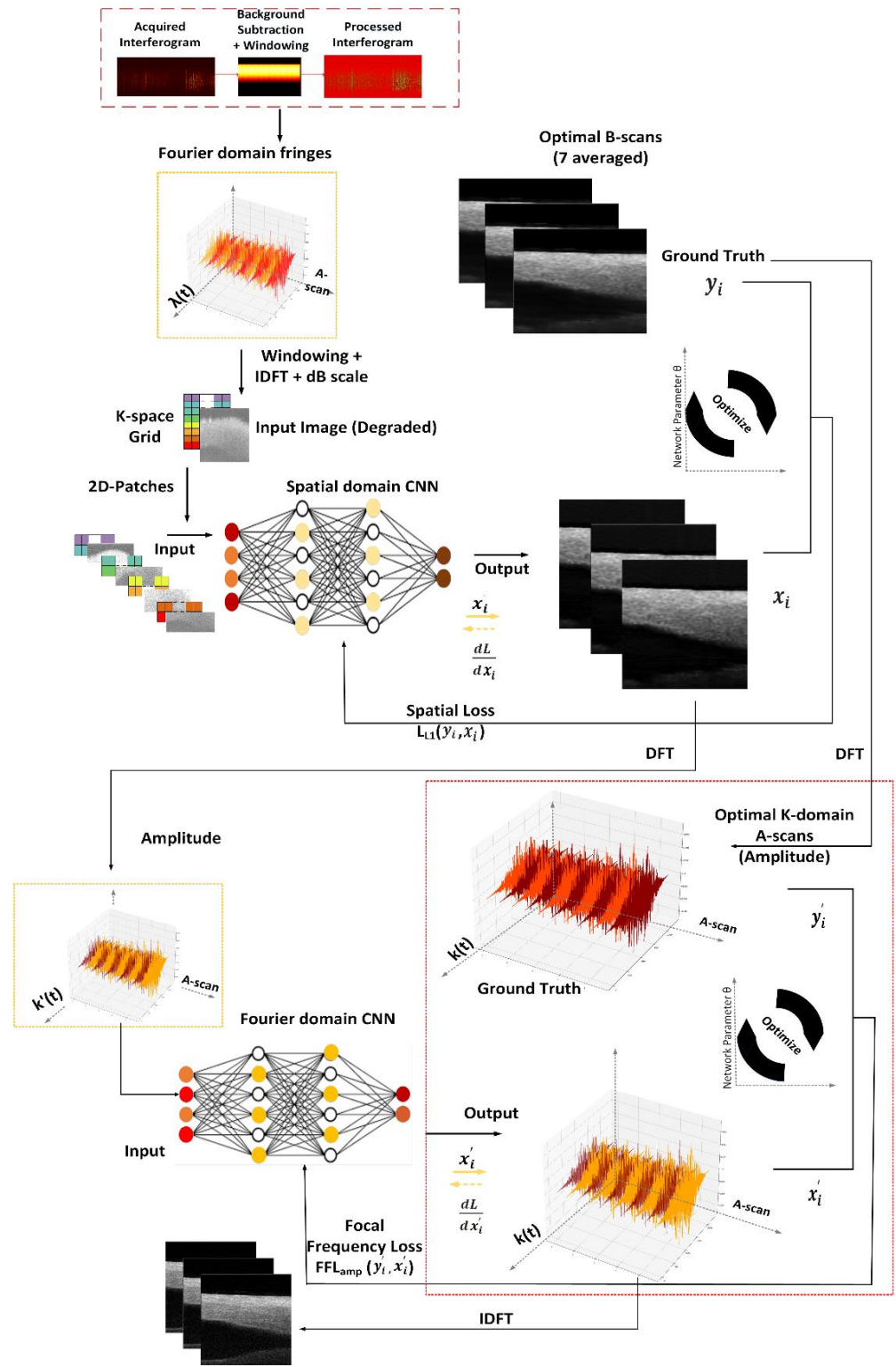

**Figure 1.** Schematics of the proposed framework containing Spatial Domain-Convolution Neural Network (SD-CNN) and Fourier Domain-Convolution Neural Network (FD-CNN), $x_i$ and $x_i'$ are the outputs and $y_i$ and $y_i'$ are the ground truths for SD-CNN and FD-CNN respectively.

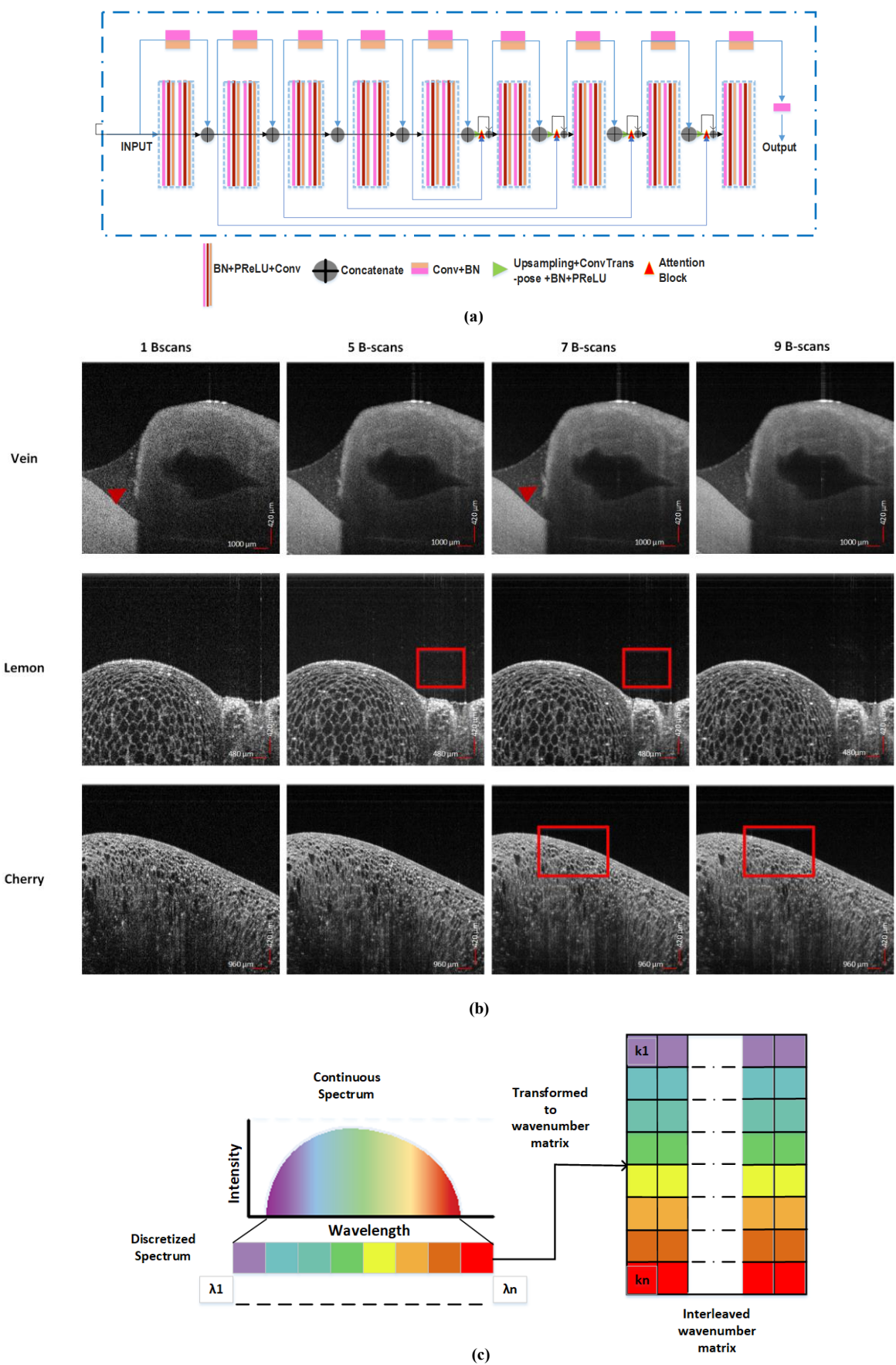

Figure 2. (a) DL network used as the Fourier Domain-Convolution Neural Network and Spatial Domain-Convolution Neural Network (b) 1 B-scan and Image averaging using 5, 7, 9 B-scans for *Vein, lemon* and *Cherry, (c)* Wavenumber layer interleaved with the input of Spatial Domain-Convolution Neural Network

In the proposed framework, we demonstrate the CNN based neural networks gov-erned by the physical law that relates the Fourier and the spatial domains. The FD-OCT systems are based on the fundamental Weiner-Khinchin theorem which states that the power spectral density P(k) (= $|I_D|^2$) of the measured signal and the auto-correlation function $\Gamma(z)$ are related by Fourier transform as follows:

$$\Gamma(z) = \int_{-\infty}^{\infty} P(k)\exp(-i2\pi kz)dk \tag{9}$$

In Eq. (9), $z$ represents the optical path lengths difference. This relation is expressed keeping in consideration the symmetrical information mirrored at positive and negative frequencies of the spectrum [35]. In a practical scenario, the obtained spectrum spans across the positive region of frequencies. This relationship is exploited by the OCT system to record data as spectrally resolved interference signal (in Fourier domain) which are then subjected to IDFT to extract the depth profile in the spatial domain. This physics-prior motivates us to design the framework for the reconstruction of OCT scans by incorporating two networks; one in the Spatial domain (SD) and other in the Fourier domain (FD). The schematics of the proposed framework is illustrated in Figure 1, depicting the workflow of the two-optimizations performed in the Spatial and Fourier domains. It is a DL-driven framework with sequential optimization of the neural networks SD-CNN and FD-CNN to reconstruct high-quality OCT images directly from the wavelength domain. The input to the SD-CNN are the spatial domain images obtained by Fourier transforming the raw interferometric data. This neural network optimizes in the spatial domain to perform high-quality image reconstruction. These images (output of SD-CNN) are further Fourier transformed and optimized in the Fourier-domain using FD-CNN for enhancing the reconstruction quality; further elaborated in sub-sections below. A diagram of the DL-model used is shown in Figure 2(a). Both networks implement this DL-model that is based on UNET as the main backbone architecture. This modified UNET consists of 4 blocks on encoder side, one block at the bottleneck followed by 4 blocks on decoder side. To boost the performance, we also use an attention gating network, residual and skip connections as shown in Figure 2(a). Attention gating helps the network by focusing on essential features required for reconstruction whereas residual connections help to address the problem of gradients in deeper networks. The encoder block is composed of Batch Normalisation (BN), activation function - PReLU, and convolution layers as stacked in the diagram shown in Figure 2(a). In addition, each block has a residual connection where input of that block is concatenated with the output of the final layer of the block. The decoder block consists of Upsampling followed by ConvTranspose, BN and PReLU layers marked with green triangle. The block output serves as one of the inputs for the attention network. A detailed description of attention block can be found in [36]. The skip connections from the encoder towards the decoder help to transfer spatial information to the decoder, but in addition these connections forward redundant low-level features which can be suppressed using the attention block. This block suppresses activation functions from regions having redundant information. To train the framework for speckle reduction, the ground truth images are obtained by averaging 7 consecutive B-scans to supress the speckle noise. This can be done due to similarity between consecutive scans. Here, we represent in Figure 2b, the images obtained from the OCT system as 1 b-scan and results of averaging 5, 7, and 9 B-scans for *vein*, *lemon* and *cherry* samples. The 1 B-scan images clearly show high speckle noise for all the three samples. Further, for the different samples we can compare and visualise the following: (i) for *vein* the improvement in the appearance of structure (marked with red arrow) (ii) the reduction of the background noise for *lemon* in 5 and 7 B-scans (marked with red box), (iii) progressive over smoothening in *cherry* sample as averaging includes more B-scans. Taking into account, the reduction of speckle noise, the over-smoothening and the lateral resolution to avoid excessive blurring, as demonstrated before in several state-of-art DL-methods for image enhancement or denoising [37]-[38], we use 7 averaged B-scans as ground truth in this work. The detailed training of the two networks along with the associated data-processing is described in sub-sections below.

*2.2. Spatial domain CNN*

The modified encoder-decoder model used in the SD-CNN, follows from the architecture described above and illustrated in Figure. 2(a) with some modifications in the network. The convolution layers in this network are 2D as this network works in the spatial domain. The input data for the SD-CNN network are prepared by processing the wavelength-domain raw spectrum. This pre-processing involves background subtraction to remove the fixed pattern noise followed by spectral shaping using Hann windowing. Then SD-CNN takes these pre-processed unevenly spaced fringes transformed into low quality images using IDFT as the input. They suffer from degraded resolution due to immense blurring arising from the non-linearity of the data in wavenumber-domain. Next, we obtain the ground truth images ($y_i$) for SD-CNN, using the commercial OCT system [30]. The ground truth images in spatial domain are obtained by averaging 7 consecutive B-scans as mentioned above, to remove the speckle noise. Furthermore, we supplemented the network with knowledge of the wavenumber range, as an additional input layer while training the SD-CNN as shown in Figure 1 (K-space grid) and elaborated in Figure 2(c). As discussed in section 2.1, $\lambda$ spans over the wavelength range ($\lambda_{min}, \lambda_{max}$), which corresponding to pixels in depth (axially). This range is sampled into $N$ (total) number of sampling points and $s$ refers to the sampling point. We calculate the non-uniformly sampled wavenumber-domain points as follows:

$$\lambda = \lambda_{min} + \frac{s}{N-1}(\lambda_{max} - \lambda_{min}) \quad (10)$$

$$k = \frac{2\pi}{\lambda} = \frac{2\pi}{\lambda_{min} + s(\lambda_{max}-\lambda_{min})/(N-1)} \quad (11)$$

These non-uniform wavenumbers are crucial as they are Fourier transform pairs with the pixel position in depth (axially) for each A-scan. Hence, to inform the network about the non-linearity of the acquired input, we add a secondary layer to the original input (1152 × 256), using $k$ values estimated from Eq. (11), as a column of matrix shown in Figure 2(c). As the same wavenumber-values in the column correspond to each A-scan, this column vector is repeated for all rows of the raw input. This interleaved k - layer in parallel to the input spectrum, serves in providing more physical-context to the network. These non-linear wavenumbers help the data-driven CNN to adjust the weights and learn accordingly. They are further divided into four 2-D patches of size 288 × 256 to enhance the learning and convergence of the model.

*2.3. Fourier domain CNN*

The FD-CNN is based on the architecture described above and shown in Figure 2(a) with minor modifications. As the A-scans have 1-D dependencies, the FD-CNN uses only 1-D convolution kernels for feature extraction. The input data for the FD-CNN network are prepared by processing the output of the SD-CNN ($x_i$) in the Fourier domain. This pre-processing involves DFT where we extract the amplitude and phase. The ground truth ($y_i'$) for the FD-CNN is the amplitude obtained by performing DFT on the ground truth images (7 averaged B-scans) originally used by the SD-CNN and described in section 2.2. This ground truth has nearly linear in wavenumber-domain spectrum as the commercial system [30] incorporates k-linearization for obtaining the final OCT images. The FD-CNN takes these amplitude values in the Fourier domain and minimizes the loss to generate the evenly spaced wavenumber-domain during the training phase. To predict the final results, the FD-CNN output $x_i'$ which is the optimized amplitude is subjected to IDFT and the phase information is utilised from the Fourier transformed output of the SD-CNN.

*2.4. Loss Function*

In the proposed framework, we use two networks with different loss functions. For the SD-CNN, we use the mean absolute error (L1) to minimize the loss between the pixels of the low-resolution (LR) image generated after applying IDFT to the linear in wavelength-domain spectrum and the high-resolution (HR) ground truth (7-averaged B-scans) obtained from the OCT system. The L1 loss is calculated as:

$$L_{L1} = \frac{1}{IJ}\sum_{i=1}^{I}\sum_{j=1}^{J} |HR - LR|, \tag{12}$$

where, i and j are the spatial indices in Eq. (12). The FD-CNN uses the Focal Frequency Loss (FFL) [32] to minimize the loss between the input $\lambda$-domain ($F_\lambda(u,v)$) and the ground truth wavenumber-domain ($F_k(u,v)$) spectra, with u and v are the indices of frequency coefficients. $F_\lambda(u,v)$ and $F_k(u,v)$ are obtained after applying DFT on the output of SD-CNN and on the ground truth images respectively. The FFL can be expressed as follows:

$$FFL(u,v) = \frac{1}{MN}\sum_{u=0}^{M-1}\sum_{v=0}^{N-1} w(u,v) |F_k(u,v) - F_\lambda(u,v)|^2 \tag{13}$$

$$w(u,v) = |F_k(u,v) - F_\lambda(u,v)|^\alpha \tag{14}$$

Here, in Eq. (13), M and N correspond to the spectrum size and in Eq. (14) w(u,v) represents the weight matrix, where the scaling factor $\alpha$ is set to 1 in this work. As the B-scan reconstruction from the raw data comprises only the amplitude of the spectrum, we minimized the error using FFL only for the amplitude using the notation $FFL_{amp}$.

## 3. Experiments

### 3.1. Imaging and Dataset processing

Optical Coherence Tomographic imaging was performed on a MHz SS-OCT benchtop system from Optores GmbH, Munich [30]. In the reported experiments, the SS-OCT uses a Fourier domain mode-locked (FDML) laser with central wavelength 1310 nm and bandwidth 100 nm. The axial resolution is 15 μm (air), lateral resolution 39.5 μm, sweeping speed is 1.6 MHz, and the lateral field of view is 10mm × 10mm. Seven different samples (or objects) namely *vein, finger, lemon, tooth, cherry, flounder egg*, and *seed (pea)* were used as OCT volume datasets. For all volumes, each B-scans had 1024 A-scans and each A-scan had 2304 points in depth for the acquired $\lambda$-space raw spectrum. For final processing, the raw-data and the corresponding B-scans (images) were cropped to a size 2304 × 256 (the raw data before IDFT) due to memory limitations. After IDFT as mentioned in pre-processing (Section 2), only half of the signal (mirror symmetry) is considered resulting into a size 1152 × 256.

The custom designed MHz-OCT software allows the user to extract data from the SS-OCT system at different stages of processing. The system captures the interferometric signal corresponding to each point in the 3D data set using the swept-source FDML laser (1310 nm). The sweeping across wavelength allows depth information to be encoded in the spectral domain. This interferometric signal is detected using dual-balanced photoreceiver and processed to obtain the depth resolved profile of the sample. The interreference signals are digitised using a fast PCIe (Peripheral Component Interconnect Express) and further linearized in the wavenumber domain. Once linear in k-domain, they are Fourier transformed to obtain the reflectivity profile. To account for the wavelength-dependent shifts in the phase, dispersion compensation is performed. The 3D data are acquired via two Galvano-scanning mirrors that scan in x and y directions in a predefined range. The depth-resolved profiles are used to generate the so-called B-scans (x-z) referred to as "OCT Output" in this work. For further details on the Optores OCT system can be referred [30].

In the context of this study, we utilize the following: (i) raw data which are the raw interference spectra in $\lambda$-space, (ii) OCT B-scans, (iii) an averaged OCT B-scan. The OCT data are standardized using the mean and the standard deviation for DL-framework. Adam is used as the optimizer with a learning rate $10^{-4}$ for both the networks. The SD-CNN converges at 200 epochs while the FD-CNN with its further fine-tuning requires around 400 epochs.

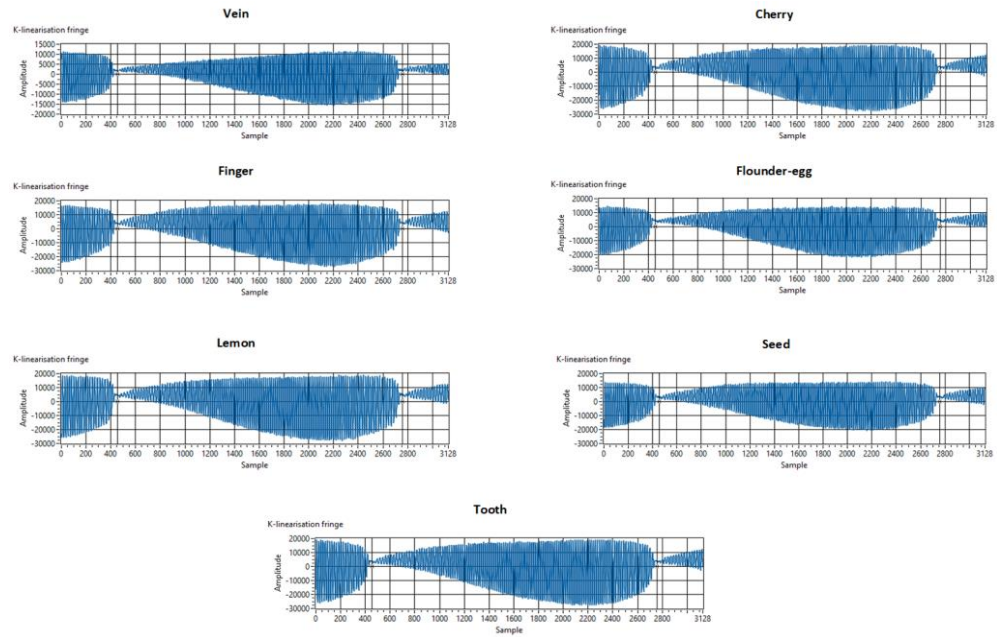

**Figure 3**. Comparison of k-linearization fringes for different volumes used: *vein, finger, lemon, tooth, cherry, flounder-egg, seed (pea)*

### 3.2. Generalizability

In order to have a robust and generalizable framework, we utilize a dataset with different acquisition parameters for field of view, calibration pattern and material properties. The field of view in the lateral (x-y) dimensions for various samples are: *lemon* (4 mm × 4 mm), *vein* (8 mm × 8 mm), *cherry* (8 mm × 8 mm), *tooth* (4 mm × 4 mm), *finger* (8 mm × 8 mm), *seed (pea)* (10 mm × 10 mm), and *flounder egg* (3 mm × 3 mm).

The k-linearization fringes for all the 7 volumes are shown in Figure 3, which demonstrate that each volume has different calibration pattern, acquired at the start of data acquisition process by the OCT system [30].

The refractive indices are important in OCT imaging as it uses a light source for creating volumetric scans of samples. In this study, various samples with differing refractive indices were used, approximately as follows: *human vein tissue* (1.3–1.4), *lemon and cherry* (1.47), *human finger skin* (1.42*), human tooth* (2.6–3.1), *seed* (pea) (1.5–1.7), and *flounder egg* (1.3–1.4). This allows to generate varying spectrum patterns for samples with different material properties, helping to train and test a more versatile model.

### 3.3. Training, Testing and Validations

The implementation was performed on a CPU AMD Ryzen 7 with random-access memory- 64GB, and the GPU has the graphic card Nvidia RTX 3090. The DL framework comprises the SD-CNN and FD-CNN, with the SD-CNN receiving low-quality B-scans as input. These B-scans are derived from raw $\lambda$-domain fringes following the pre-processing steps, as described in Section 2.2. The output upon Fourier-domain transformation is supplied to the FD-CNN, which optimizes the network in the Fourier-domain. This FD-CNN output is subjected to IDFT to get final B-scans as described in Section 2.3. Regarding the training of these networks, the SD-CNN is independently trained first and then the FD-CNN is trained by freezing the weights of the SD-CNN with its inputs and processing described in Sections 2.2 and 2.3. Similarly, in the inference stage, first the SD-CNN takes degraded input image obtained from the wavenumber domain along with k-space grid in a fashion as shown in Figure 1 for the training and then DFT is performed on the output of SD-CNN and the amplitude is fed into the FD-CNN to predict the fringes, which are then subjected to IDFT to obtain the resultant OCT images. The single DL- framework uses the two models sequentially to infer the output.

From the 7 volumes described above, 5 volumes namely *lemon, vein, cherry, tooth, finger* containing 3000 B-scans (600 in each volume) were used for training, validation, and testing. These 3000 B-scans were partitioned randomly into training, validation, and testing groups with percentages 70, 20, and 10 for the DL-framework proposed. Each one of the other two volumes, namely *seed (pea)* and *flounder egg*, contains 200 B-scans (size 1152 × 256). These B-scans were reserved to test the generalization capability of the proposed deep learning method described in cross-validation (Section 3.4.2). Thus, no information from these two volumes was used in the training or validation procedure.

*3.4. Results*

As discussed in Section 2, the SD-CNN and FD-CNN were trained sequentially, using two different loss functions namely L1 loss and $FFL_{amp}$. After their individual trainings, they were tested to infer the reconstructions also in a sequential manner, in the same order as they were trained.

*3.4.1. Performance evaluation of the entire frame work*

We use different quantitative metrics namely MSE (Mean Square Error), PSNR (Peak Signal to Noise Ratio), SSIM (Structural Similarity Index Measure), CNR (Contrast to Noise Ratio) for evaluating the reconstruction results. CNR is the measure of contrast when comparing the foreground region with the signal and the background region affected by noise in the image, calculated over the i$^{th}$ regions using mean $\mu_i$ (foreground), mean $\mu_b$ (background) and standard deviations $\sigma_i$ (foreground) and $\sigma_b$ background respectively as:

$$CNR_i = 10 log_{10}\left(\frac{|\mu_i - \mu_b|}{\sqrt{\sigma_i^2 + \sigma_b^2}}\right), \quad (15)$$

In addition, we also use another metric, namely $\beta_s$ that measures of the degree of smoothness in images. Using representation as $\mu$ (mean), $I$ (2D-image with *x* and *y* as indices of pixels), *out* (output) and *in* (input), $\beta_s$ can be calculated as:

$$\beta_s = \frac{\Gamma(I_{out} - \mu_{out}, I_{in} - \mu_{in})}{\sqrt{\Gamma(I_{out} - \mu_{out}, I_{out} - \mu_{out}) \cdot \Gamma(I_{in} - \mu_{in}, I_{in} - \mu_{in})}}, \quad (16)$$

$$\text{where } \Gamma(I_1, I_2) = \sum_{x,y}[I_1(x,y) \cdot I_2(x,y)]. \quad (17)$$

We discuss in this section the performance of the proposed model and compare the results between the degraded input, OCT output (generated by the Optores OCT system [30]) and high-quality ground truth. The evaluation is done on the segregated randomly selected test dataset, which was not exposed to the models during their training stage. In Figure 4 we evaluate the performance of the framework. The images obtained from the raw data linear in $\lambda$ are fed to the SD-CNN followed by Fourier domain transformation on the output (of SD-CNN) whose amplitude is optimized using the FD-CNN. We compare the ground truth (A), the OCT output (B), the degraded raw data input to the SD-CNN (C), and the final output (D) from the proposed framework in Figure 4. It represents images from each of the 5 samples used in this work. Compared with the degraded inputs, the final outputs show well-reconstructed images retaining structures similar to the ground truth images. The proposed framework effectively produces enhanced quality compared to the commercial OCT output (B) which is obtained without any averaging. In Figure 4, the *vein*, the *tooth* and the *finger* which are characterized by larger uniform regions show clear representation of reduced speckle noise, especially when compared to the images from OCT output.

The quantitative evaluation of the performance is done using the metrics namely SSIM, PSNR, $\beta_s$ and CNR to compare the inference of the proposed framework with the degraded input and the OCT output shown in Table 1 (with best results highlighted). The test dataset is used to calculate PSNR, SSIM and $\beta_s$ parameter using the ground truth images as the reference. The averaged overall PSNR shows an improvement of approximately 2dB and 13 dB respectively for the OCT Output and Input (degraded input to SD-CNN) compared to the output of the proposed framework. It is to be noted that we get a fairly high PSNR, especially for the *lemon* and *cherry* datasets, that comprise highly structured images. In contrast, the tooth reconstruction from the input (degraded OCT image) provides a gain of 12 dB, although it remains a little below the output generated by the OCT system that is 22.11dB.

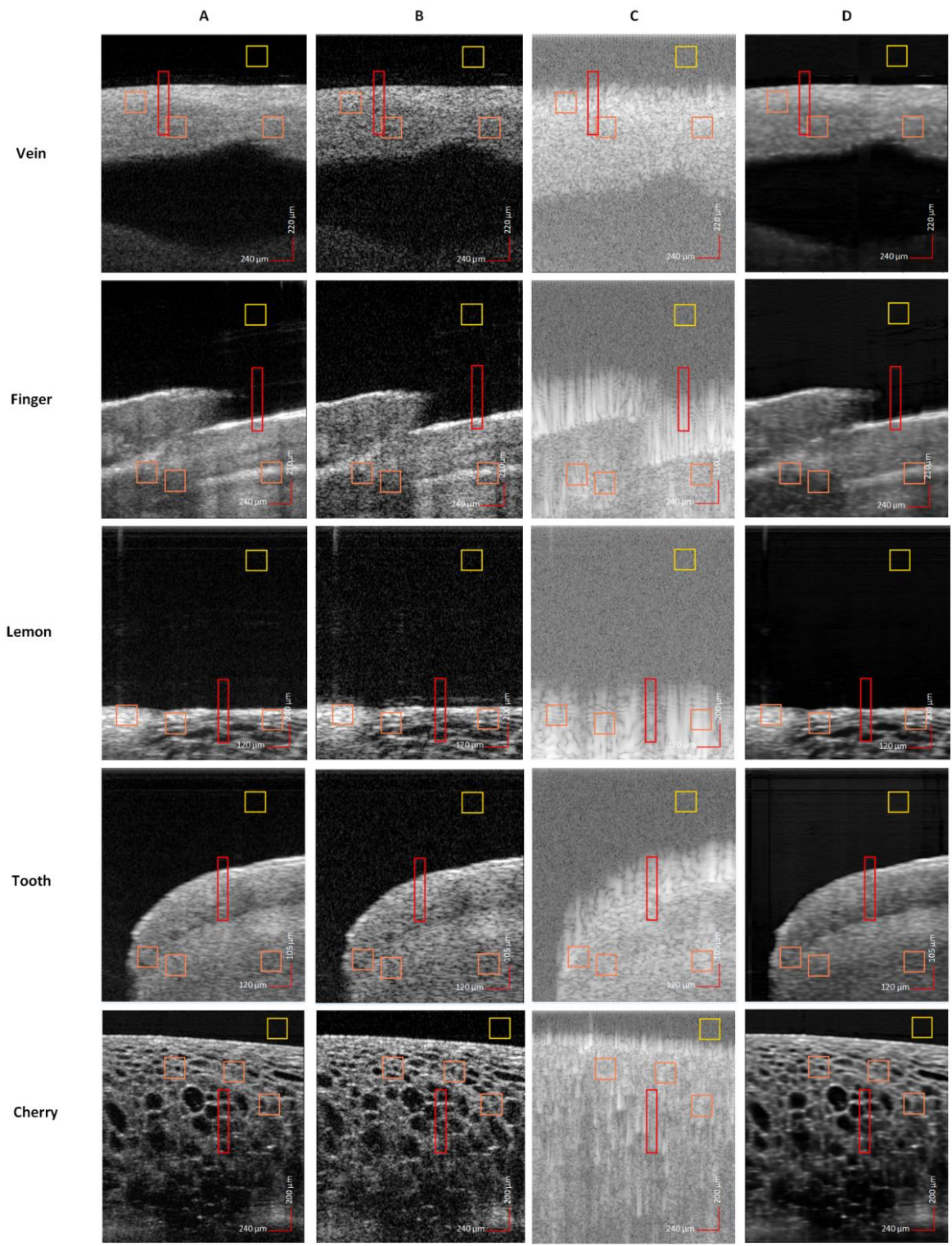

**Figure 4.** Comparison of B-scans from five different volumes. (A) Ground truth (seven B-scans of OCT system output averaged), (B) OCT system output, (C) OCT system raw data input, (D) output of the proposed framework. The reconstruction (D) shows high similarity to the desired ground truth.

Using SSIM with the ground truth as a reference, higher degree of similarity can be seen with the final output. Though the SSIM values show increment and are precise, they

**Table 1.** Comparison of scores for PSNR, SSIM, CNR, $\beta$ parameter for Input- degraded input to SD-CNN, OCT Output (generated by the Optores OCT system), and reconstructed OCT images from the proposed framework

| Method | Dataset | PSNR | SSIM | CNR | $\beta$ |
|---|---|---|---|---|---|
| Input | Overall | 8.94 | 0.08 | - | 0.71 |
| | Vein | 12.76 | 0.14 | 4.44 | 0.68 |
| | Finger | 7.92 | 0.06 | 4.62 | 0.75 |
| | Lemon | 7.14 | 0.05 | 5.69 | 0.64 |
| | Tooth | 10.11 | 0.12 | 4.31 | 0.77 |
| | Cherry | 8.79 | 0.07 | 5.04 | 0.73 |
| OCT Output | Overall | 19.95 | 0.35 | - | 0.87 |
| | Vein | 21.20 | 0.22 | 5.21 | 0.88 |
| | Finger | 19.93 | **0.45** | 3.04 | 0.92 |
| | Lemon | 20.40 | 0.26 | 6.73 | 0.78 |
| | Tooth | **22.11** | **0.56** | 0.75 | **0.93** |
| | Cherry | 17.55 | 0.30 | 3.96 | 0.85 |
| Proposed | Overall | **22.30** | **0.46** | - | **0.93** |
| | Vein | **21.98** | **0.43** | **8.71** | **0.93** |
| | Finger | **21.75** | 0.42 | **4.86** | **0.94** |
| | Lemon | **25.74** | **0.54** | **7.98** | **0.91** |
| | Tooth | 21.61 | 0.32 | **6.48** | 0.92 |
| | Cherry | **21.67** | **0.62** | **4.86** | **0.95** |

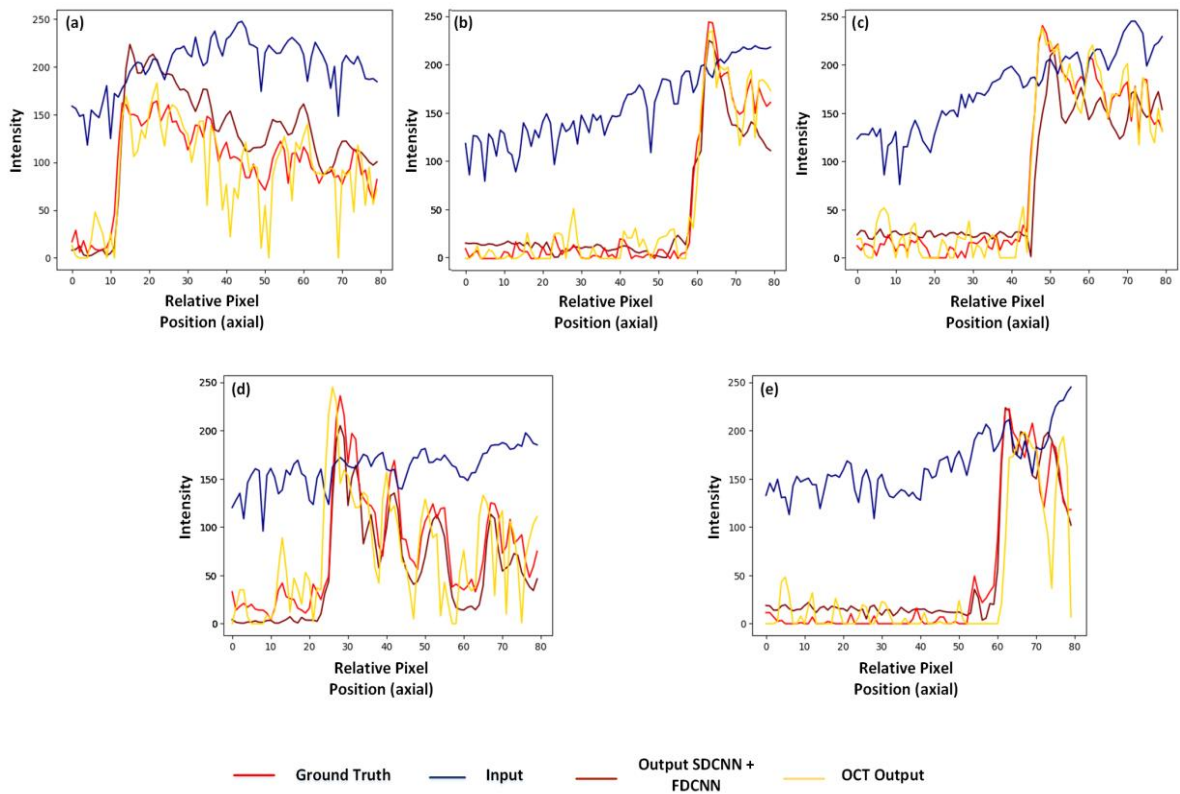

**Figure 5.** Comparison of line plot to show the variation of intensity (A.U.) for the central column of red rectangle marked in Figure 4, here plots correspond to (a*) vein*, (b) *inger* (c) lemon, (d) *tooth* and (e) *cherry samples*, for ground truth, input OCT output and proposed framework outputs shown using different lines in each plot

might not directly reflect image quality if used independently in the OCT assessments

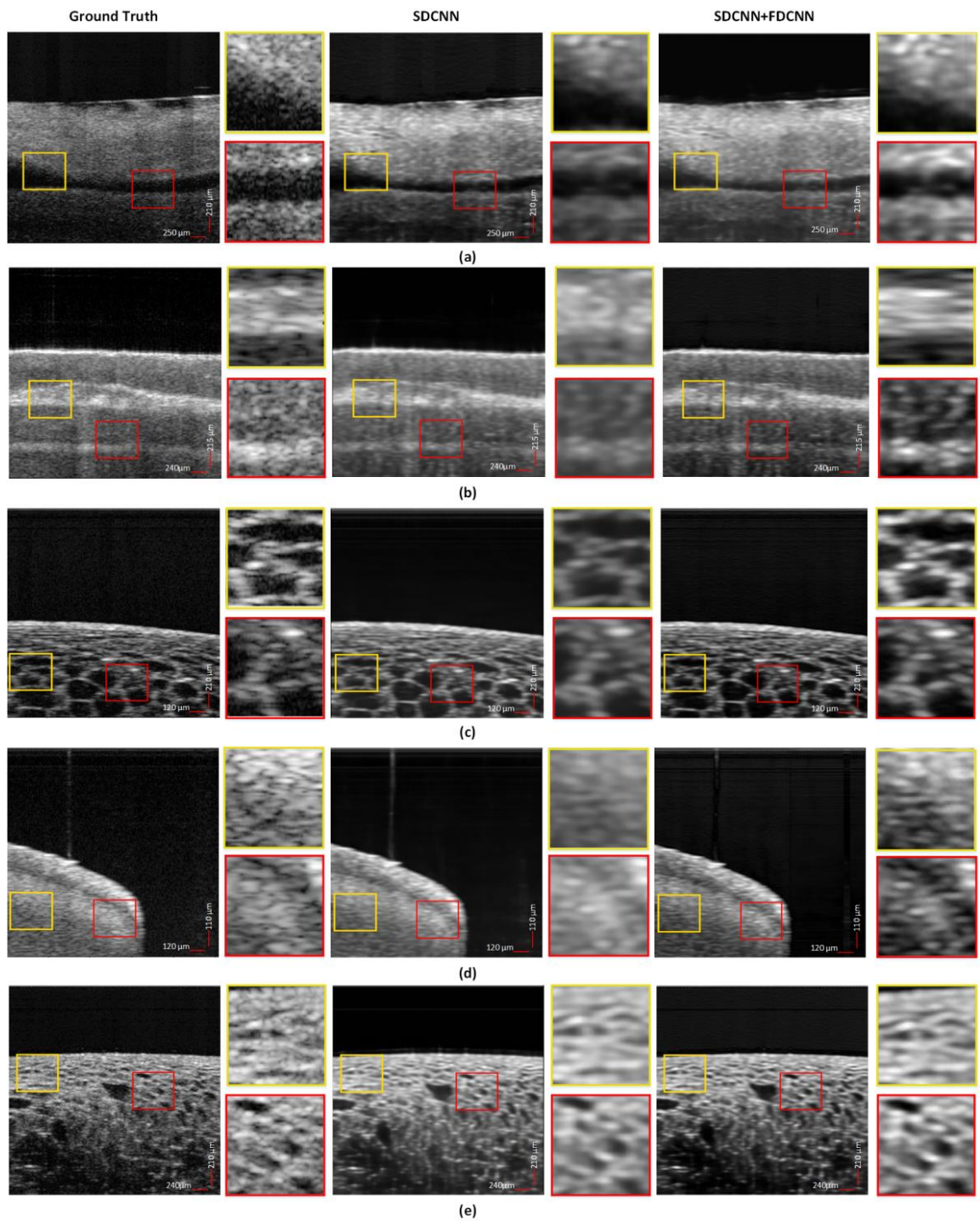

**Figure 6**. Comparison between the ground truth, the output of the Spatial Domain-Convolution Neural Network and Fourier Domain-Convolution Neural Network for (a) *vein*, (b) *finger*, (c) *lemon*, (d) *tooth* and (e) *cherry*. For each image, the magnified regions are shown for better comparison. The results of the combined SD-CNN+FD-CNN show enhanced performance when compared to output of only SD-CNN, demonstrating the better reconstruction capability of high-frequency details using FD-CNN.

[39]. The granular speckle noise may be misinterpreted as structures resulting in inaccurate assessments while calculating SSIM scores. The $\beta_s$ parameter (Eq. (16)) helps to determine how well the enhanced image preserves the structural features along with reducing speckle noise in the OCT images. There is an overall improvement from 0.87 to 0.93 between the OCT system output and the proposed DL-framework's output, indicating significant reduction of speckle noise in the reconstructed images.

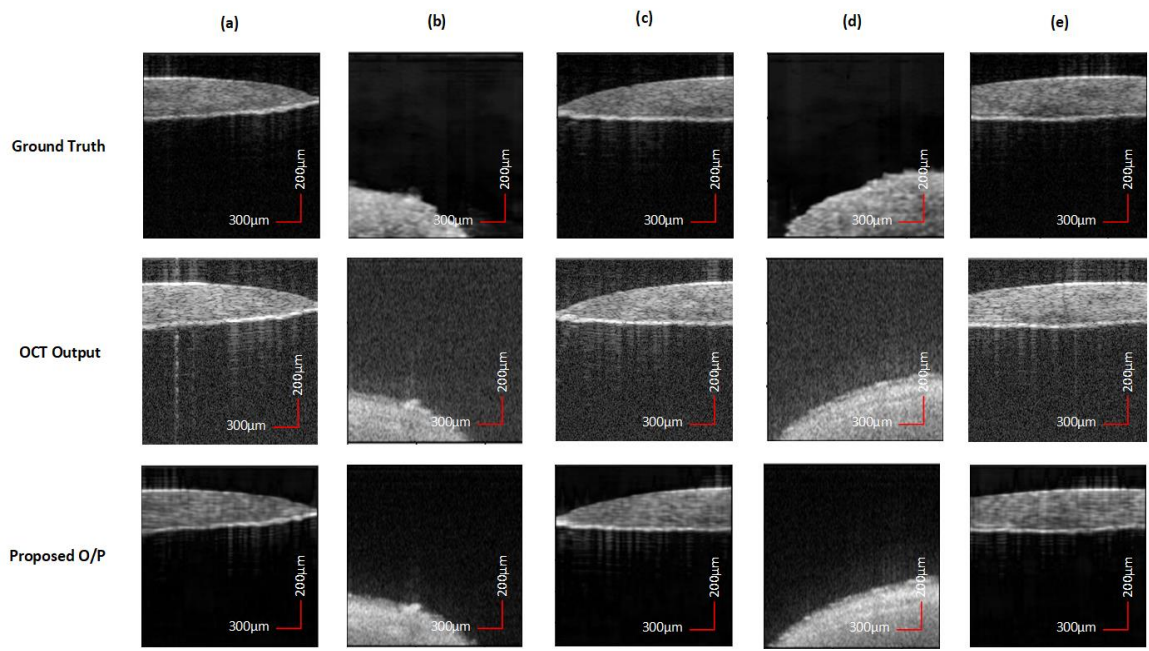

**Figure 7.** Comparison of cross-validation results on (a),(c),(e) *flounder egg* and (b),(d) *seed (pea)* samples for ground truth, OCT output and reconstructions obtained by the proposed model. The cross-validation results show robustness and generalization capability of the proposed model on completely unseen volume.

**Table 2.** Cross-validation results

| Volume | PSNR | SSIM |
| --- | --- | --- |
| *Flounder egg* | 22.09 | 0.45 |
| *Seed (pea)* | 21.53 | 0.42 |

We also examined the CNR over the regions marked by the orange boxes in Figure 4 using Eq. (15) with $i = 3$ for the foreground regions and the yellow box for the background region. The findings indicate that the proposed model is capable to depict higher contrast of the crucial features for an image in comparison to noise irrespective of the sample type.

In an attempt to monitor the variations at boundaries (or edges), we analyse the column (pixels) at the central position of the red rectangular region in Figure 4 by plotting the intensity variations in Figure 5 for all the five types of sample images (input, ground truth, OCT output and final output). Here the x-axis is relative pixel position, where the first row of the rectangle is referred as 1 and is not indicative of the exact pixel position (exact) as in the original image. The ground truth is marked red and our output in maroon closely follows it when compared to the OCT output which is afflicted by numerous spikes due to presence of speckle even in the smooth regions. The input as expected follows divergent trajectory due to degradations caused by blur (as an outcome of non-linear wavenumber-based spectrum) and noise.

The performances of both models SD-CNN and FD-CNN are compared in Figure 6, where we show the inferred samples from the test dataset along with respective ground truths. The five rows contain images from each volume namely *vein*, *finger*, *lemon*, *tooth* and *cherry* (for row 1-5 respectively). Each image contains two regions marked with yellow and red boxes that are further magnified (~ 5X) as adjacent images, for better visualisation. The examples show that the SD-CNN can reconstruct well when the image contains more-uniform areas whereas the FD-CNN provides boost to the over-all framework by enhancing the structural details and features. In particular, for the *lemon* and the *cherry* we can visualize the clear manifestation of the morphological structures. The output of combined SD-CNN and FD-CNN models can suppress the speckle noise and significantly improve the structures and boundaries. The homogeneity attributed to the output images using the two models is highly desirable in OCT application as it enhances the image quality greatly. This superior generalization capability stems from the use of the physics prior which optimizes both Fourier and spatial domains.

**Table 3.** Ablation study to compare effects of Fourier Domain-Convolution Neural Network (FD-CNN) and Spatial Domain-Convolution Neural Network (SD-CNN) on reconstruction quality

| SD-CNN | FD-CNN | SD-CNN+FD-CNN | PSNR | | SSIM | |
|---|---|---|---|---|---|---|
| | | | Avg | Std | Avg | Std |
| ✓ | - | - | 20.81 | 2.64 | 0.42 | 0.11 |
| - | ✓ | - | 10.97 | 0.80 | 0.03 | 0.01 |
| - | - | ✓ | 22.30 | 2.51 | 0.46 | 0.08 |

**Table 4**. Time comparison

| Stepwise Time Complexity (s) | |
|---|---|
| **Operations** | **Time (s)** |
| Calibration | 0.008 |
| Resampling in K-domain | 0.17 |
| FFT | 0.05 |
| Averaging (Speckle reduction) | 0.07 |

| Overall time Complexity -Volume (s) | | | |
|---|---|---|---|
| *OCT system [30]* | *Fourier Domain-CNN* | *Spatial Domain-CNN* | *Fourier Domain-CNN + Spatial Domain-CNN* |
| 792 | 68.55 | 79.723 | 142.5 |

*3.4.2. Cross-validation and ablation studies*

We perform cross-validation studies in Table 2 on 2 different volumes namely (i) *seed (pea)* (ii) *flounder egg* each containing 200 images to assess the results on completely unseen volumes which were not shown to the framework during training or validation. These samples have different texture and material properties compared to the volumes used in training or validation of the model. We tabulate average value of SSIM and PSNR scores obtained on the two volumes. In addition, we can also visualise the results in Figure 7 and compare the reconstruction performed by the model, the OCT output from the system and the ground truth images. The cross-validation performance of the proposed DL-framework shows robust reliability and generalization capability of the model on completely new data.

We conduct ablation study to assess how each of the FD-CNN and SD-CNN network contributes to the overall performance. The study is presented in Table 3 using PSNR and SSIM having average and standard deviation values. The SD-CNN, when trained and inferred individually, provides an image quality having average PSNR and SSIM equals to 20.81dB and 0.42 respectively. In contrast, when only the FD-CNN is trained and inferred in a similar individual manner, the model performs poorly with average PSNR of 10.97dB and SSIM 0.03 . The CNN exhibit superior performance for images in the spatial domain, compared to only-Fourier domain, owing to their inherent architecture, receptive field, and hierarchical feature extraction. When we evaluate the combined approach (the proposed framework), we get increased outputs for both PSNR and SSIM with average values of 22.30 dB and 0.46 and standard deviation of 2.51dB and 0.08 respectively.

*3.4.3. Time-complexity*

In Table 4, we provide the time-complexity for several processing steps like calibration, resampling and FFT for one B-scan and averaging done to remove speckle noise on 7 consecutive B-scans. The implementation was performed on a CPU AMD Ryzen 7 with random-access memory- 64GB, and the GPU has the graphic card Nvidia RTX 3090 as mentioned previously in Section 3.3. It is to be noted that these are only few important processing steps performed to obtain the single B-scan from raw data. The commercial systems like [30] perform several other operations to handle different types of artifacts such as noise, DC term, dispersion, etc. It can be seen that for 1 B-scan, resampling takes quite a substantial amount of time i.e. 0.17 seconds. Also, this single B-scan is highly affected by speckle in practical scenarios, and methods like averaging (either adjacent scans or registration method performed on several volumes) required to reduce speckle, requires multiple B-scans. Hence much more time is taken to produce these multiple scans for reducing speckle compared to generating individual B-scan afflicted by speckle noise.

We compare the time complexities for reconstructing a volume of 600 images (each of size 1152 × 1024) using the proposed model versus the commercial OCT system [30]. The latter was employed for raw data acquisition and to generate ground-truth images with speckle reduction, achieved by averaging 7 B-scans. The proposed model ensures 142.5s computation time for this volume whereas the commercial system takes 792s. Individual implementations comprising of only FD-CNN and the SD-CNN takes 68.55s and 79.723s respectively. The OCT system [30] processes (including averaging) multiple B-scans to reduce the speckle noise, and generate smooth B-scan. This increases the computational complexity. In contrast, the proposed method does not depend on multiple B-scan processing for generating speckle reduced B-scan which makes it quite faster. The faster processing allows handling large volumes of OCT data with more optimized workflow in medical and laboratory settings. It enables comprehensive integration of OCT system with advanced technological set-ups for timely analyses.

## 4. Discussion

OCT based image reconstruction problem for nonlinear in wavenumber domain data, has been an active research topic for decades now. The widespread applications of this technology in several bio-medical domains, demand for faster image processing techniques along with high fidelity images. The FD-OCT systems having acquired spectrum linear in wavelength domain should undergo calibration and k-mapping prior to performing IDFT to extract the depth profile for OCT images. In this work, we propose an approach based on the DL-framework to offer a paradigm reform by reconstructing OCT B-scans directly from their acquired raw wavelength domain spectrums. The two neural networks incorporated in this framework effectively model the respective representations in spatial and Fourier domain. They together achieve OCT reconstructions retaining morphological details with reduced speckle noise. Furthermore, the demonstrated computational efficiency is also critical for time-sensitive applications that particularly require fast image reconstruction. Thus, the proposed work not only focuses on reconstruction of high-fidelity images but also reduces the computational complexity significantly. Such improvements, besides catering to present day OCT applications, will in-addition open room for innovative, faster-processed speckle reduced images that are crucial for integration with advanced technologies, portability, cost-effectiveness and efficient use of resources. One of the limitations of this work is the comparison with only one OCT system by Optores [30]. The hardware and software complexity associated with different commercial OCT systems and the unavailability of procedural details makes it difficult to reproduce the exact pipeline involved from raw data acquisition to final image reconstruction. This constrained setup requirement precluded the comparisons with other independent systems involving different stages of the workflow. In future, we aim to pursue avenues to extend presented work for different OCT systems and settings.

and E.S.; supervision, E.R., V.M., and E.S.; project administration, V.M. and E.S.; funding acquisition, V.M. and E.S. All authors have read and agreed to the published version of the manuscript."

**Funding:** This research was funded by European Union's Horizon 2020 research and innovation programme under the Marie Skłodowska Curie grant agreement No 956770.

**Data Availability Statement**: Dataset available on request from the authors.

**Acknowledgments:** M.V. would like to thank European Union's Horizon 2020 research and innovation programme under the Marie Skłodowska-Curie grant agreement No 956770 for the funding. V.M. and E.S. would like to thank European Regional Development Fund within the Operational Programme "Science and Education for Smart Growth 2014–2020" under the Project CoE "National center of Mechatronics and Clean Technologies" BG05M2OP001-1.001-0008. E.R. would like to acknowledge the support by the Academy of Finland (project no. 336357, PROFI 6 - TAU Imaging Research Platform).

**Conflicts of Interest:** "The authors declare no conflicts of interest."